\documentclass[12pt]{elsarticle}

\usepackage{lipsum}
\makeatletter
\def\ps@pprintTitle{%
 \let\@oddhead\@empty
 \let\@evenhead\@empty
 \def\@oddfoot{}%
 \let\@evenfoot\@oddfoot}
\makeatother

\usepackage{subcaption}
\usepackage[utf8]{inputenc}
\usepackage[export]{adjustbox}
\usepackage{wrapfig}
\usepackage{graphicx}
\usepackage{amssymb}
\usepackage{amsmath, nccmath}
\usepackage{commath}
\usepackage{amsthm}
\usepackage{geometry}
\usepackage{lineno}

\usepackage{lipsum}
\usepackage{array}
\makeatletter
\def\ps@pprintTitle{%
 \let\@oddhead\@empty
 \let\@evenhead\@empty
 \def\@oddfoot{}%
 \let\@evenfoot\@oddfoot}
\makeatother

\begin{document}

\begin{frontmatter}


\title{Continued functions and Borel-Leroy transformation: \\ Resummation of six-loop $\epsilon$-expansions from different universality classes}



\author{Venkat Abhignan, R.  Sankaranarayanan}

\address{Department of Physics, National Institute of Technology, Tiruchirappali - 620015, India. }

\begin{abstract}
We handle divergent $\epsilon$ expansions in different universality classes derived from modified Landau–Wilson Hamiltonian. Landau–Wilson Hamiltonian can cater for describing critical phenomena on a wide range of physical systems which differ in symmetry conditions and the associated universality class. Numerically critical parameters are the most interesting physical quantities which characterize the singular behaviour around the critical point. More precise estimates are obtained for these critical parameters than previous predictions from Pad\'e based methods and Borel with conformal mapping procedure. We use simple methods based on continued functions and Borel-Leroy transformation to achieve this. These accurate results are helpful in strengthening existing conclusions in different $\phi^4$ models.

\end{abstract}


\end{frontmatter}

\section{Introduction}
The calculation of recent six-loop \cite{six-O(n)} and seven-loop  renormalization functions \cite{seven-O(n)} in $O(n)$-symmetric $\phi^4$ field theory has lead to significant improvement in results related to universal critical phenomenon \cite{shalaby2020critical,shalaby2020}, compared to the 25 years old five-loop functions \cite{Guida_1998,legui}. It has been helpful in solving for accurate universal parameters and also better understand the decade old discrepancy between theoretical  predictions and experimental value in $O(2)$ $\phi^4$ model \cite{lambda}. This was achieved using Hypergeometric-Meijer resummation to solve for critical exponents \cite{shalaby2020critical,shalaby2020}. The issue commonly called as "$\lambda$-point specific heat experimental anomaly" was also addressed by using a new blended, continued exponential fraction \cite{abhignan2020continued}. We solved for the critical exponents of arbitrary $n$-component field in different spatial dimensions with the help of other continued functions such as continued exponential and continued fraction  in the $O(n)$-vector model. It is a scalar field theory which has an infrared fixed point and around its vicinity the system occupies symmetries of homogeneous, isotropic and dilation in nature. This makes it a favourable ground to study different physical systems at phase transitions with similar underlying behaviour but leading to different universality classes. $O(1)$ $\phi^4$ theory can analyse phase transitions on Ising model, simple fluids and binary mixtures. $O(2)$ $\phi^4$ theory  deals with superconductivity and superfluid helium-4 transition. Similarly transitions on Heisenberg ferromagnets ($n=3$), some models of quark-gluon plasma ($n=4$), neutron star matter ($n=10$) and superfluid helium-3 ($n=18$) can be studied. \\ 
Quantities derived from field-theoretical perturbative methods are typically divergent and resummation procedures are needed to extract meaningful estimates with high precision. Resummation techniques have been the most common way to handle perturbation series from field theories. With rapid development in computational methods the sixth order \cite{six-O(n)} and seventh order \cite{seven-O(n)} terms in such series were obtained quite progressively and soon can lead to more orders. This demands the need for accessible resummation methods which provide accurate results with only lower order information. Traditionally resummation techniques such as Pad\'e approximants and Pad\'e-Borel-Leroy transformation are used for the purpose of obtaining convergence by only using the lower order information of the perturbation series \cite{ADZHEMYAN2019,KOMPANIETS2020,RIM2021}. Continued fractions are equivalent to Pad\'e approximants, while they can be also related to Shanks transformation which are used to accelerate the convergence nature \cite{Shanks}. The modified orthogonal hypergeometric approximants used in Hypergeometric-Meijer resummation can also be represented from a particular form of continued function, namely Gauss’s continued fractions. Such continuous iterative functions have been used to tackle divergent series in different ways \cite{Yukalov2019}, which is an old idea developed by Yukalov \cite{Yukalov1991,Yukalov1992}. Here we implement resummation methods using continued functions and Shanks transformation on recently derived six-loop $\epsilon$ expansions in different modified models. Further to improve convergence and remove irregularities we combine continued functions with Borel-Leroy transformation. Such six-loop perturbative expansions were derived in $n$-vector model with cubic anisotropy \cite{ADZHEMYAN2019}, $O(n)\times O(m)$ spin models \cite{KOMPANIETS2020} and weakly disordered Ising model \cite{RIM2021}, which were all solved based on the six-loop expansions from $O(n)$-symmetric model by Kompaniets et al. \cite{six-O(n)}. Re-estimation of critical parameters in these models using different methods with better accuracy are helpful in removing contradictions and strengthening existing arguments regarding these theories. Also in all these cases the exact solutions are not known and the reliability of the extrapolated estimates is only established when different methods produce comparable results. \\ The paper is organized as follows: We initially introduce the continued functions and their implementation through Borel-Leroy transformation in Sec. 2. We then handle the perturbative expansions from $n$-vector model with cubic anisotropy, $O(n)\times O(m)$ spin models and weakly disordered Ising model in Sec. 3, 4 and 5 respectively.
\section{Continued functions and Borel-Leroy transformation}
The $\epsilon$ expansions in all the models produce quantities of interest in form of \begin{equation}
    Q(\epsilon)\approx\sum^N_{i=0} q_i \epsilon^i
\end{equation}  in a system with $d=4-\epsilon$ spatial dimensions. As mentioned earlier this quantity is divergent in nature and hence we convert $Q(\epsilon)$ into continued exponential (CE),  \begin{equation}
    Q(\epsilon) \sim b_0\exp(b_1\epsilon \exp(b_2 \epsilon \exp(b_3 \epsilon \exp(b_4 \epsilon\exp(b_5 \epsilon\exp(b_6 \epsilon\exp(\cdots)))))))
\end{equation} or a continued exponential fraction (CEF), \begin{equation}
   Q(\epsilon) \sim
 c_0\exp\left(\frac{1}{1+c_1\epsilon\exp\left(\frac{1}{1+c_2\epsilon\exp\left(\frac{1}{1+c_3\epsilon\exp\left(\frac{1}{1+\cdots}\right)}\right)}\right)}\right) \; \; \; \hbox{for} \; \; \; (\epsilon \rightarrow 0). 
\end{equation} Coefficients $\{b_i\}$ and $\{c_i\}$ can be solved up to arbitary order $i$ from the perturbative coefficients $\{q_i\}$ by Taylor expansion of CE and CEF as shown previously \cite{abhignan2020continued}. These affine transformations generally enlarge the region of convergence for $Q(\epsilon)$ by a different mapping. And the convergence is obtained by observing the successive approximants \begin{equation}
    B_1 \equiv b_0\exp(b_1\epsilon),\,B_2 \equiv b_0\exp(b_1\epsilon\exp(b_2\epsilon)),\,B_3 \equiv b_0\exp(b_1\epsilon\exp(b_2\epsilon\exp(b_3\epsilon))),\,\cdots
\end{equation}
in case of CE and the sequence 
\begin{multline}
     C_1 \equiv c_0\exp\left(\frac{1}{1+c_1\epsilon}\right),\,C_2 \equiv c_0\exp\left(\frac{1}{1+c_1\epsilon\exp\left(\frac{1}{1+c_2\epsilon}\right)}\right),\\ C_3 \equiv c_0\exp\left(\frac{1}{1+c_1\epsilon\exp\left(\frac{1}{1+c_2\epsilon\exp\left(\frac{1}{1+c_3\epsilon}\right)}\right)}\right),\cdots
 \end{multline} in case of CEF. Such a sequence is further converged with aid of Shanks transformation and its iteration for any sequence $\{A_i\}$ such as, \begin{equation}
  S(A_i) = \frac{A_{i+1}A_{i-1}-A_i^2}{A_{i+1}+A_{i-1}-2A_i}\,\,\,\hbox{and}\,\,\,S^2(A_i) = \frac{S(A_{i+1})S(A_{i-1})-S(A_i^2)}{S(A_{i+1})+S(A_{i-1})-2S(A_i)}.
 \end{equation} \\
 Further the Borel-Leroy (BL) technique we use is derivative of Pad\'e-BL transformation commonly used  \cite{ADZHEMYAN2019,KOMPANIETS2020,RIM2021}, where the Pad\'e approximant is replaced with CE. BL transformation uses a shift parameter $l$ based on which we can optimize the convergence behaviour. Tuning this parameter $l$ we can tame the factorial growth of coefficients in quantity $Q(\epsilon)$ and refine the estimate of $Q(\epsilon)$ to higher accuracy. The BL transformation is given by \begin{equation}
     Q(\epsilon) = \int_0^\infty e^{-t} t^l F(\epsilon t) dt ,\,\,\, F(y) = \sum_{i=0}^\infty \frac{q_i}{\Gamma(i+l+1)} y^i.
 \end{equation} This transformation which implements sequence of CE is given by \begin{multline}
     D_1 \equiv \int_0^\infty e^{-t} t^l d_0\exp(d_1\epsilon t) dt,\,D_2 \equiv \int_0^\infty e^{-t} t^l d_0\exp(d_1\epsilon t\exp(d_2\epsilon t)) dt,\\ D_3 \equiv \int_0^\infty e^{-t} t^l d_0\exp(d_1\epsilon t\exp(d_2\epsilon t\exp(d_3\epsilon t))) dt,\cdots
 \end{multline} where coefficients $\{d_i\}$ are obtained similar to $\{b_i\}$ from the series $F(y)$. Similar to previous approaches the sequence $\{D_i\}$ provides the converged estimate which is further accelerated using Shanks transformation. We observe that implementing CEF with BL transformation does not lead to significant improvement in the convergence behaviour. \\ To measure the accuracy of our estimates we calculate error using the relation \begin{equation}
     (|S(A_{i+1}) - S(A_{i})| + |S(A_{i+1}) - S^2(A_{i})|)/2,
 \end{equation} when we take $S^2(A_{i})$ as our final estimate for $Q(\epsilon)$ from the sequence $\{A_i\}$. This relation which is based on successive approximations gives the accuracy of our estimate since the value of Shanks iteration depends on previous iterations. To find the parameter $l$ in BL transformation, we inspect the space of $l \in [0,50]$ with $\Delta l=0.01$ to minimize the error given above. We choose the resummation tool empirically based on observing the convergence nature of coefficients $\{q_i\}$ and feasibility of finding parameter $l$. \\ We have already checked the applicability of CE and CEF on perturbative $\epsilon$ expansions from $O(n)$-symmetric field theory \cite{abhignan2020continued}. To test the applicability of CE with BL transformation, we implement it to calculate the correction-to-scaling exponent $\omega$ from $O(n)$-symmetric theory. Especially in $3d$ system for the $O(4)$ $\phi^4$ model we previously used continued fraction to obtain the estimate $\omega=0.7896(1)$ \cite{abhignan2020continued}, which is only compatible with recent calculations from self-consistent resummation algorithm \cite{sc2020} where $\omega=0.7863(9)$ was obtained. It is also comparable with estimates from Borel with conformal mapping (BCM) and conformal bootstrap calculations, $\omega=0.794(9)$ \cite{six-O(n)} and $\omega=0.817(30)$ \cite{Echeverri2016}, respectively. But there is a mismatch from other recent estimates namely hypergeometric-Meijer resummation \cite{shalaby2020critical}, non-perturbative renormalization group \cite{nprg2020} and Monte Carlo simulations (MC) \cite{Hasenbusch_2001} predict as $\omega=0.7519(13)$, $\omega=0.761(12)$ and $\omega=0.765(30)$, respectively. We used continued fraction since the $\epsilon$ expansion for $\omega$ is of the form $\epsilon Q(\epsilon)$  such as \cite{shalaby2020critical} \begin{equation}
         \omega = \epsilon - 0.541667\epsilon^2 + 1.15259\epsilon^3 - 3.27193\epsilon^4 + 10.8016\epsilon^5 - 40.5665\epsilon^6 + 166.256\epsilon^7.
     \end{equation}Here we solve for $Q(\epsilon)$ using CE-BL and multiply the estimate with value of $\epsilon$. In order to obtain critical exponents for three-dimensional systems we finally equate $\epsilon$ to unity. To introduce the tuning parameter $l$ we take into consideration Eq. (7) for $Q(\epsilon)$ such as \begin{equation}
         F(\epsilon t) = \frac{1}{\Gamma(l+1)} - \frac{0.541667\epsilon t}{\Gamma(l+2)} + \frac{1.15259\epsilon^2 t^2}{\Gamma(l+3)}  - \frac{3.27193\epsilon^3 t^3}{\Gamma(l+4)}+ \frac{10.8016\epsilon^4 t^4}{\Gamma(l+5)} - \frac{40.5665\epsilon^5 t^5}{\Gamma(l+6)}+ \frac{166.256\epsilon^6 t^6}{\Gamma(l+7)}.
     \end{equation} Further by implementing CE as in Eq. (8) we find the sequence for $\epsilon Q(\epsilon)$ such as \scriptsize \begin{multline}
      D_1 = \epsilon \int_0^\infty e^{-t} t^l \frac{1}{{\Gamma }\left(l+1\right)}\exp\left(-\frac{0.541667\,{\Gamma }\left(l+1\right)}{{\Gamma }\left(l+2\right)}\epsilon t\right) dt,\\  D_2 = \epsilon \int_0^\infty e^{-t} t^l \frac{1}{{\Gamma }\left(l+1\right)}\exp\left(-\frac{0.541667\,{\Gamma }\left(l+1\right)}{{\Gamma }\left(l+2\right)}\epsilon t\exp\left(1.84615\,{\Gamma }\left(l+2\right)\,{\left(\frac{.00001\,{\Gamma }\left(l+1\right)}{{{\Gamma }\left(l+2\right)}^2 }-\frac{0.00006}{{\Gamma }\left(l+3\right)}\right)}\epsilon t\right)\right) dt,\cdots
 \end{multline}
 \small
 Evaluating these integrals we get approximants $\{D_i\}$ and further accelerate their convergence using Shanks transformed sequence $\{S(D_i)\}$. We take our final estimate for $\omega$ from $S^2(D_4)$ and estimate the error from relation $(|S(D_{5}) - S(D_{4})| + |S(D_{5}) - S^2(D_{4})|)/2$ as in Eq.(9). By tuning the parameter $l$ we illustrate the behaviour of estimate in Fig. 1(a) for $l \in [0.5,1.2]$. We observe that by shifting the parameter $l$ the nonuniformity in our approximants can be removed and region of precise estimate can be decided, where the value of estimate is sensistive to $l$. Further observing the confined region $l \in [1,1.2]$ in Fig. 1(b) we decide $l=1.097$ to get the precise estimate $\omega=0.79419$ which is most compatible with BCM prediction. We note here that in BCM procedure it is necessary to know the asymptotic behaviour of perturbative expansions for Green's functions whereas our procedure involves only lower order information.
 \begin{figure}[ht]
\centering
\begin{subfigure}{0.4\textwidth}
\includegraphics[width=1\linewidth, height=5cm]{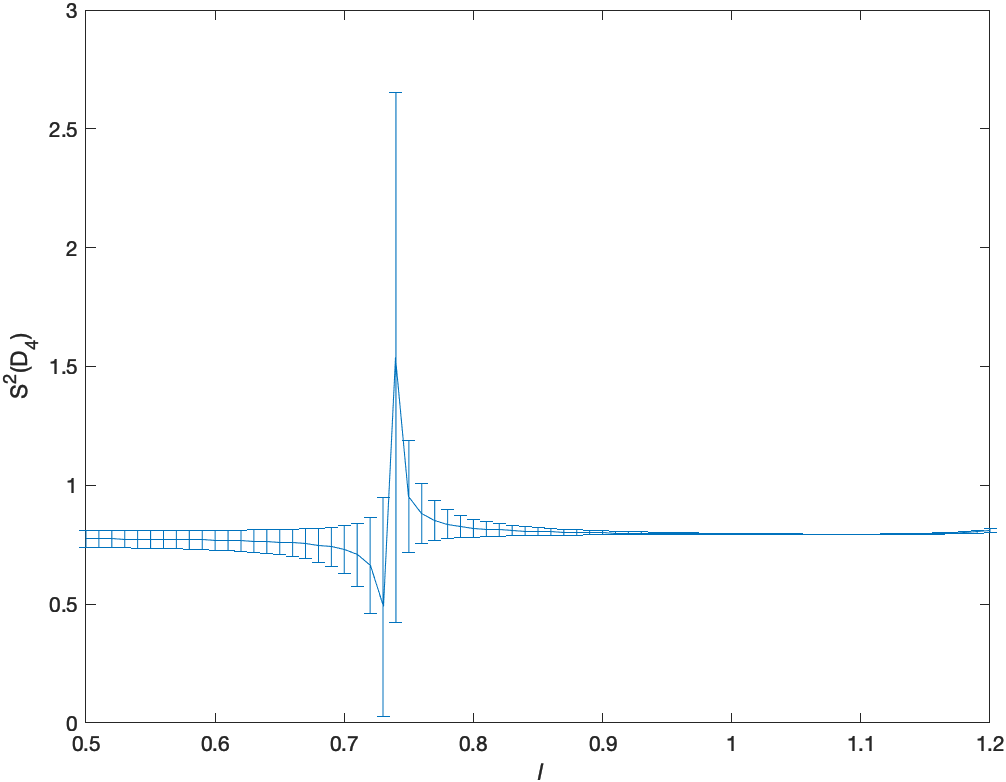} 
\caption{$S^2(D_4)$ vs $l$ for $l \in [0.5,1.2]$}

\end{subfigure}
\begin{subfigure}{0.4\textwidth}
\includegraphics[width=1\linewidth, height=5cm]{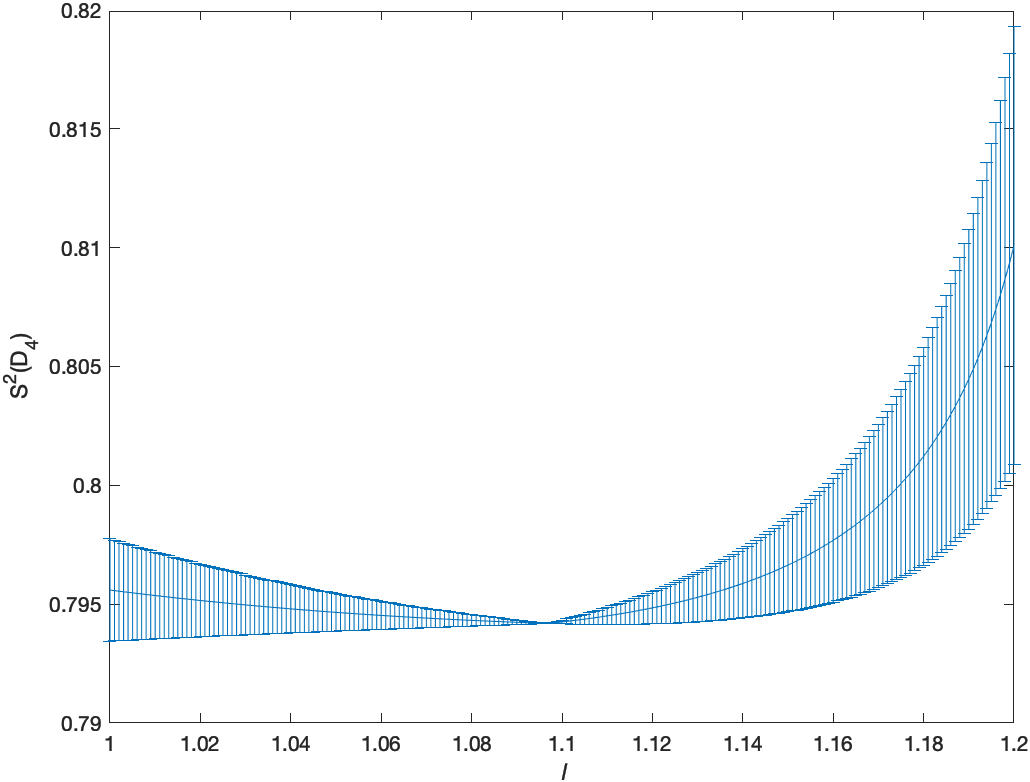}
\caption{$S^2(D_4)$ vs $l$ for $l \in [1,1.2]$}

\end{subfigure}
 
\caption{We plot the estimate of $\omega$ derived from $S^2(D_4)$ vs shift parameter $l$, with the error bars showing the value of $(|S(D_{5}) - S(D_{4})| + |S(D_{5}) - S^2(D_{4})|)/2$.}

\end{figure}
\\
The choices in this CE-BL procedure might be leading to overestimation of accuracy in the estimates obtained. More stringent test for this procedure would be to calculate correlation length exponent of well known 3d Ising model which has a general consensus up to third decimal place, $\nu_{Ising}\approx0.630$ using different theoretical approaches like renormalization group (RG) \cite{shalaby2020critical,abhignan2020continued}, Conformal bootstrap \cite{Kos2016} and MC \cite{MCHAS}. Using CE-BL resummation of slowly converging $\epsilon$ expansion \cite{shalaby2020critical} we obtain the estimate $\nu_{Ising}=0.63134$, which assures the accuracy of this procedure to at least third decimal place. \section{$n$-vector model with cubic anisotropy}
Critical parameters in a realistic cubic crystal with anisotropy were recently studied for $n$-vector field \cite{ADZHEMYAN2019}. In such model there is a competition of RG flows between isotropic Heisenberg and anisotropic cubic modes which are in different regimes of critical behaviour. This leads to stability of fixed points for this model only in a particular region of order parameter $n$. Marginal order parameter dimensionality $n_c$ determines the stability, where in case of $n<n_c$ the Heisenberg critical behaviour dominates and for $n>n_c$ cubic critical regime is stable. So for $n>n_c$, a new class of universality for critical behaviour of cubic-anisotropy emerges completely different from the $O(n)$ class (Heisenberg). Hence studying $n_c$ is of physical importance for real ferromagnets with anisotropies to determine which regime they follow. Initial studies on such models lead to $n_c>3$ using RG method \cite{maier} and $n_c=3$ using Monte Carlo simulations (MC) \cite{MC1998}. However later studies based on higher order resummation of RG perturbative series showed that $n_c<3$ \cite{PRB2000-4,PRB2000-5,PRB2000-6}. The recent six-loop $\epsilon$ expansion provides the best prediction as $n_c=2.915(3)$ with the Pad\'e-BL technique \cite{ADZHEMYAN2019}. This prediction was derived from six-loop $\epsilon$ expression for $n_c$ which was given as \begin{equation}
    n_c = 4-2\epsilon+2.588476\epsilon^2-5.874312\epsilon^3+16.82704\epsilon^4-56.62195\epsilon^5 + O(\epsilon^6)
\end{equation} for the physically interesting case of $n=3$. We convert this into CE (Eq. 2) and calculate its Shanks to get the sequence for $d=3$ as \begin{equation}
    B_1 = 2.42612,\,B_2 = 3.35454,\,B_3 = 2.63588,\,B_4 = 3.16593, B_5 = 2.71339
\end{equation} and \begin{equation}
    S(B_2) = 2.94945,\, S(B_3) = 2.94093, S(B_4) = 2.92181.
\end{equation} As we observe the sequence $\{B_i\}$ is oscillating between the best prediction given above, while the sequence of Shanks $\{S(B_i)\}$ is converging monotonically. To accelerate this convergence behaviour we take BL transformation of CE (Eq. 8) with $l=0.61$ to obtain \begin{equation}
    D_1 = 2.58794,\,D_2 = 3.18263,\,D_3 = 2.94130,\,D_4 = 2.91015, D_5 = 2.96437
\end{equation} and Shanks of the above sequence is given by\begin{equation}
    S(D_2) = 2.96437,\, S(D_3) = 2.91741, S(D_4) = 2.91735.
\end{equation} We take the final estimate from $S^2(D_3)$, $n_c=2.91735(3)$ which is more precise compared to previous estimates from 3d RG approaches \cite{PRB2000-4,PRB2000-5,PRB2000-6}, pseudo-$\epsilon$ expansion \cite{pseudoe2000,pseudoe2016} and most compatible with the recent Pad\'e-BL estimate (Table 1). We illustrate the behaviour of shift parameter $l$ versus $S^2(D_3)$ in Fig.2, with the error bars. 
\begin{figure}[ht]
\centering
\begin{subfigure}{0.4\textwidth}
\includegraphics[width=1\linewidth, height=5cm]{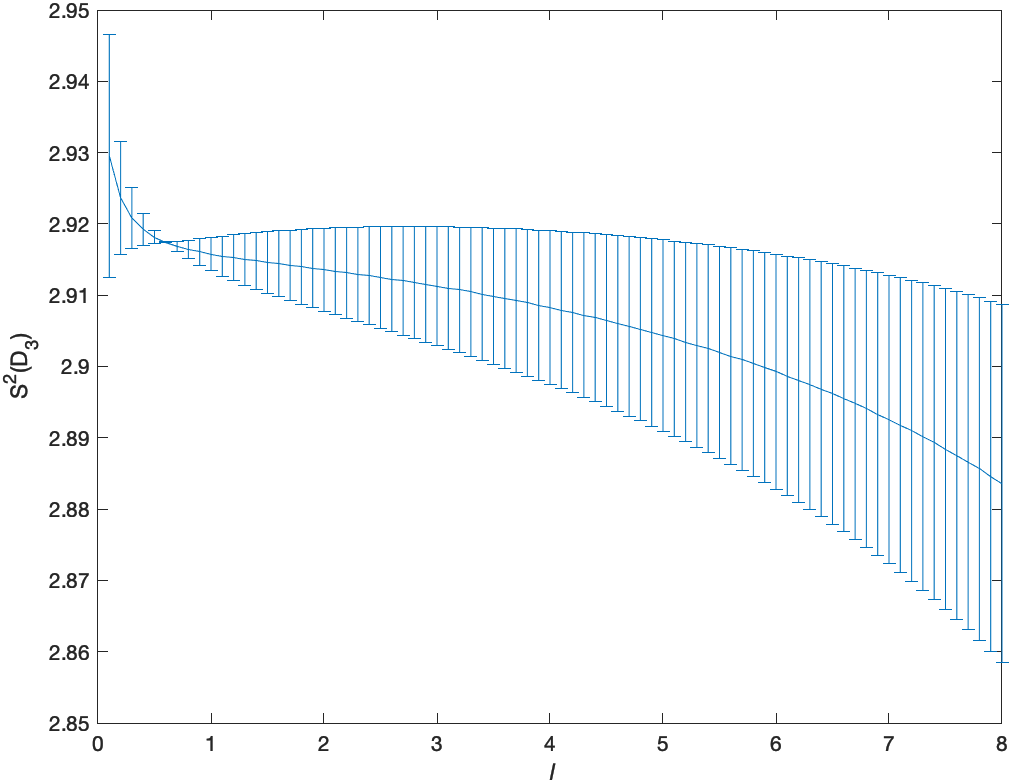} 
\caption{$S^2(D_3)$ vs $l$ for $l \in [0.1,8]$}

\end{subfigure}
\begin{subfigure}{0.4\textwidth}
\includegraphics[width=1\linewidth, height=5cm]{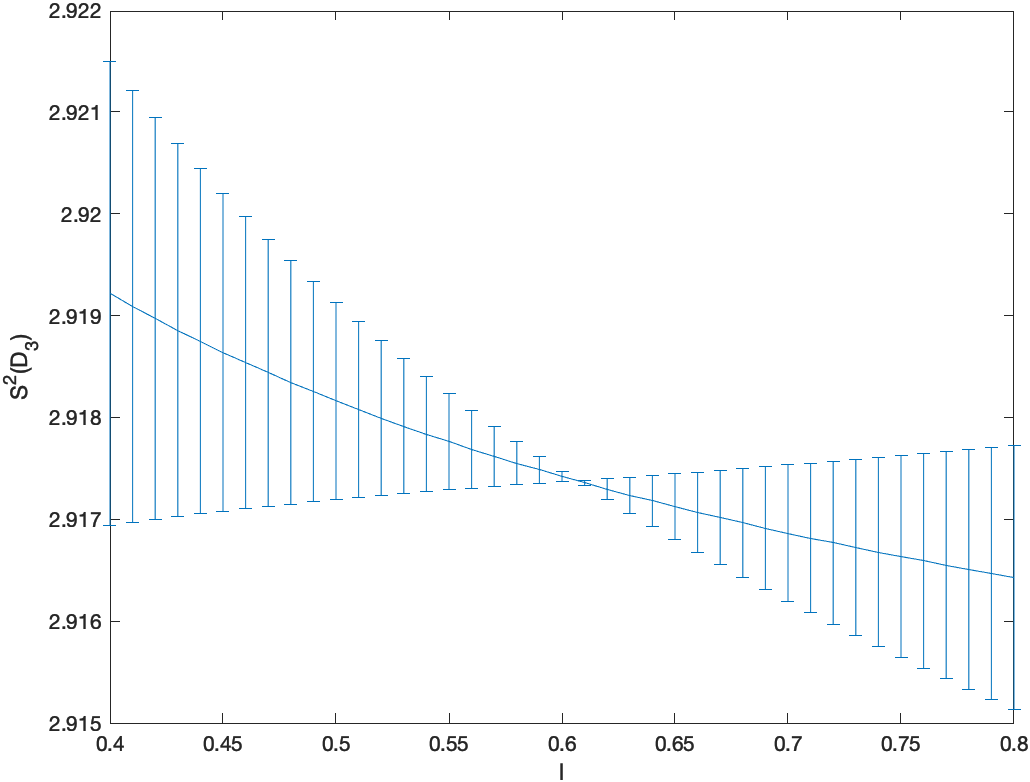}
\caption{$S^2(D_3)$ vs $l$ for $l \in [0.4,0.8]$}

\end{subfigure}

\caption{We plot the value of marginal order parameter dimensionality $n_c$ derived from CE-BL estimate $S^2(D_3)$ vs shift parameter $l$, with the error bars showing the value of $(|S(D_{4}) - S(D_{3})| + |S(D_{4}) - S^2(D_{3})|)/2$.}

\end{figure}\\ Further we study the critical exponents for this universality class of cubic ferromagnets with anisotropy  $(n=3)$, which are interesting since they differ from the Heisenberg class. The exponent $\nu$ which determines the behaviour of correlation length was given by expression \cite{ADZHEMYAN2019}
\begin{equation}
    \frac{1}{\nu} = 2-0.44444\epsilon-0.17513\epsilon^2+0.13460\epsilon^3-0.34969\epsilon^4+0.99461\epsilon^5-3.48637\epsilon^6+O(\epsilon^7).
\end{equation}
Similarly we convert this into CE for $d=3$ and obtain the oscillating sequence \begin{multline}
    1/B_1=0.62442,\,1/B_2=0.72262,\,1/B_3=0.66793,\,1/B_4=0.71540,\,1/B_5=0.68772,\\1/B_6=0.71368. 
\end{multline} Consecutively the Shanks iterations provide sequence \begin{equation}
    S(1/B_2)=0.68749,\,S(1/B_3)=0.69334,\,S(1/B_4)=0.69791,\,S(1/B_5)=0.70112,
\end{equation}
and
\begin{equation}
    S^2(1/B_3)=0.714305,\,S^2(1/B_4)=0.70858.
\end{equation}
We observe that the final iterated values of Shanks $S^2(1/B_3)$ and $S^2(1/B_4)$ provide a good neighbourhood for the prediction of $\nu$. To obtain a more precise prediction, we convert the expression of $\nu$ into CEF (Eq. 3) whose rapidly convergent sequence is \begin{equation}
    1/C_2=0.63725,\,1/C_3=0.77547,\,1/C_4=0.70693,\,1/C_5=0.71451,\,1/C_6=0.71122,
\end{equation} 
with corresponding Shanks as
\begin{equation}
    S(1/C_3)=0.72965,\,S(1/C_4)=0.71376,\,S(1/C_5)=0.71222,\,S^2(1/C_4)=0.71205.
\end{equation}
Our precise CEF estimate for cubic class $\nu=0.71205(85)$ is better than previous Pad\'e-BL value of $\nu=0.700(8)$ \cite{ADZHEMYAN2019}. And it marginally differs from recent seven-loop precise estimates of Heisenberg class using CEF, $\nu=0.70787(39)$ \cite{abhignan2020continued} and hypergeometric-Meijer resummation, $\nu=0.70906(18)$ \cite{shalaby2020critical}. Similarly using the expression for $\gamma$ \cite{ADZHEMYAN2019} the critical exponent of susceptibility, we determine its CE estimate using $S^2(B_4)=1.404(16)$ and CEF estimate using $S^2(C_4)=1.4236(11)$. Similarly to calculate the correction-to-scaling exponent $\omega$ we take the series which is of the form $\epsilon Q(\epsilon)$ \cite{ADZHEMYAN2019}, solve for $Q(\epsilon)$ using CE and CE-BL and multiply the estimate with value of $\epsilon$. Further we compare them all with previous most compatible cubic class estimates and also with recent estimates from Heisenberg class in Table 1. These precise estimates show that quantitatively the critical exponents of 3d cubic class and Heisenberg class are numerically close and differ very marginally. This difference may be refined with help from higher $\epsilon$ expansions.
\begingroup
\setlength{\tabcolsep}{6pt} 
\renewcommand{\arraystretch}{1} 
\begin{table}[htp]
\scriptsize
\begin{center}
\caption{Critical parameters $n_c$, $\nu$, $\gamma$, $\omega$ for class of $n=3$ cubic-anisortropy model in three dimensional systems} 

 \begin{tabular}{||c c c c||}
 
 \hline
Critical parameter & Continued function estimates & Existing predictions & \begin{tabular}{c c}
     &  Existing predictions \\
     & (Heisenberg class)
\end{tabular} \\ [0.5ex] 
 \hline\hline
 
     $n_c$
   
   & \begin{tabular}{c c c c}
   & 2.956(26) (CE) \\
        & 2.91735(3) (CE-BL)
   \end{tabular} 
   & \begin{tabular}{c c c c c}
        & 2.915(3) \cite{ADZHEMYAN2019}\\
        & 2.89-2.92 \cite{PRB2000-5}  \\ 
        & 2.89(4) \cite{PRB2000-6}\\
        & 2.862(5) \cite{pseudoe2000} \\
        & 2.86(1) \cite{pseudoe2016} 
        \end{tabular}
        
        & --- \\
    
 \hline
     $\nu$
   & \begin{tabular}{c c c c}
    & 0.7085(53) (CE) \\
        & 0.71205(85) (CEF)
   \end{tabular} 
   & \begin{tabular}{c c}
        & 0.700(8) \cite{ADZHEMYAN2019} \\
        & 0.706(6) \cite{PRB2000-4}\\
        & 0.704(4) \cite{PRB2000-6}
   \end{tabular}
   & \begin{tabular}{c c}
             & 0.70787(39) (CEF) \cite{abhignan2020continued} \\
             & 0.70906(18) \cite{shalaby2020critical}
        \end{tabular} \\
 \hline
    $\gamma$
   
   & \begin{tabular}{c c c c}
   & 1.404(16) (CE) \\
        & 1.4236(11) (CEF) 
   \end{tabular} 
   & \begin{tabular}{c c}
        & 1.368(12) \cite{ADZHEMYAN2019} \\
        & 1.419(6) \cite{PRB2000-6} \\
        & 1.416(4) \cite{pseudoe2000} 
   \end{tabular}
   & \begin{tabular}{c c}
             &  1.3929(46) (CE) \cite{abhignan2020continued}\\
             & 1.385(4) \cite{ADZHEMYAN2019}
        \end{tabular} \\
 \hline
     $\omega$
   
   & \begin{tabular}{c c c c}
   & 0.78417(29) (CE) \\
        & 0.78569(68) (CE-BL)
   \end{tabular} 
   & \begin{tabular}{c c}
        & 0.799(4) \cite{ADZHEMYAN2019} \\
        & 0.7833(54) \cite{PRB2000-4} \\
        & 0.781(4) \cite{PRB2000-6} 
   \end{tabular} & \begin{tabular}{c c}
             &  0.79083(1) \cite{abhignan2020continued}\\
             & 0.794(4) \cite{sc2020}
        \end{tabular} \\ 
 \hline
     
\end{tabular}
\label{table 1}
\end{center}
\end{table}
\section{$O(n)\times O(m)$ spin models}
The extension of $O(n)$-symmetric field theory to $O(n)\times O(m)$ symmetry is helpful in studying frustrated spin systems with noncoplanar and noncollinear ordering \cite{KOMPANIETS2020}. Such spin structures can be found in physical systems such as stacked triangular antiferromagnets (STA) and helical magnets (HM). Such physically interesting systems can be studied in this case for $n=2$, XY and $n=3$, Heisenberg frustrated antiferromagnets. For $m=2$ the theory corresponds to system with noncollinear but coplanar ordering (such that $n\geq m$). Further with $m\geq3$ we can study the critical behavior of magnets with noncoplanar ordering. Similar to the previous model based on stability of fixed points in RG flows, we can deduce the existence of an upper marginal order parameter value $n^+(m,d)$ and lower marginal dimensionalities  $n^H(m,d)$, $n^-(m,d)$. This demarcation of order parameter $n$ is useful to differentiate the different regimes of critical behaviour (where $n^H(m,d)<n^-(m,d)<n^+(m,d)$). For $n>n^+(m,d)$ the system follows second-order continuous phase transition with "chiral" universality class and for $n^-(m,d)<n<n^+(m,d)$ the transition is first-order. Further for $n<n^H(m,d)$ Heisenberg or $O(mn)$-symmetric critical behaviour dominates and for $n^H(m,d)<n<n^-(m,d)$ the system either undergoes first-order transition or ordering happens with simple unfrustrated sinusoidal spin structure depending on the coupling constant. Finding these parameters along with critical exponents for such chiral universality class had given contradictory results in field-theoretical models compared to experiments \cite{PhysRevB.63.140414,PhysRevB.65.020403}. Also the results of different experiments had produced a wide range of unrelated critical exponents and even the order of phase transition was found to differ for materials with same symmetry \cite{PRL-FRUST, PhysRevB.67.094434, PhysRevB.62.8983, PhysRevB.66.052405, PhysRevB.66.184432, PhysRevB.67.104431, PhysRevB.93.024412, PhysRevB.96.224427, PhysRevB.99.094408}. \\
Hence initially getting precise estimates for upper marginal dimensionality  $n^+(m=\{2,3\},4-\epsilon)$ is of physical importance to determine the type of phase transition in real systems. We illustrate the results for determining $n^+(m=\{2,3\},4-\epsilon)$ which seem to possess the most irregular numerical structure compared to other quantities of lower marginal dimensionalities \cite{KOMPANIETS2020}. The six-loop $\epsilon$ expansion for $n^+(2,4-\epsilon)$ was obtained as \cite{KOMPANIETS2020} \begin{equation}
    n^+(2, 4 - \epsilon) = 21.798 - 23.431 \epsilon + 7.0882 \epsilon^2 - 0.0321 \epsilon^3 + 4.2650 \epsilon^4 - 8.4436 \epsilon^5 + O(\epsilon^6).
\end{equation}
Constructing CE and consecutive Shanks iteration of this expansion for $d=3$ we obtain \begin{multline}
    B_1 = 7.440,B_2 = 5.597,B_3 = 5.323,B_4 = 5.596, B_5 = 5.597 \\ \hbox{and}\,\, S(B_2) = 5.275, S(B_3) = 5.450, S(B_4) = 5.597, S^2(B_3) = 5.996.
\end{multline}
Similarly constructing CE-BL ($l=3.69$, Fig 3.(a)) and consecutive Shanks iteration we obtain \begin{multline}
    D_1 = 8.281,D_2 = 6.402,D_3 = 5.911,D_4 = 6.147, D_5 = 6.034 \\ \hbox{and}\,\,
    S(D_2) = 5.737, S(D_3) = 6.07, S(D_4) = 6.07, S^2(D_3) = 6.07.
\end{multline} We observe our estimates from CE, $n^+(2,3)=5.99(27)$ and CE-BL, $n^+(2,3)=6.07$ are compatible with recent inverse biased Pad\'e-BL prediction $n^+(2,3)=6.1(1)$ and BCM prediction $n^+(2,3)=6.0(6)$ \cite{KOMPANIETS2020}. Similarly the six-loop $\epsilon$ expansion for $n^+(3,4-\epsilon)$ was obtained as \cite{KOMPANIETS2020} \begin{equation}
 n^+(3,4-\epsilon) =  32.492 - 33.719 \epsilon + 11.100 \epsilon^2 - 2.1440 \epsilon^3 + 5.2756 \epsilon^4 - 8.4830 \epsilon^5 + O(\epsilon^6). 
\end{equation} Constructing CE and consecutively Shanks iteration of this expansion for $d=3$ we obtain \begin{multline}
    B_1 = 11.5102,B_2 = 9.267,B_3 = 8.408,B_4 = 9.177, B_5 = 9.109 \\ \hbox{and}\,\, S(B_2) = 7.876, S(B_3) = 8.814, S(B_4) = 9.115, S^2(B_3) = 9.257. 
\end{multline} Constructing CE-BL ($l=5.27$, Fig 3.(b)) and consecutively Shanks iteration we obtain \begin{multline}
    D_1 = 12.437,D_2 = 9.989,D_3 = 9.084,D_4 = 9.532, D_5 = 9.310 \\ \hbox{and}\,\,
    S(D_2) = 8.554, S(D_3) = 9.384, S(D_4) = 9.384, S^2(D_3) = 9.384.
\end{multline} We observe our estimates from CE, $n^+(3,3)=9.25(22)$ and CE-BL, $n^+(3,3)=9.384$ are compatible with the inverse biased Pad\'e-BL prediction $n^+(3,3)=9.7(1.0)$ and BCM prediction $n^+(3,3)=9.3(4)$ \cite{KOMPANIETS2020}. Similarly using the CE and CE-BL methods we compute estimates of other physically interesting values of upper marginal order parameter $n^+(m=\{4,5,6\},4-\epsilon)$ and lower marginal dimensionalities  $n^-(m=\{2,...,6\},4-\epsilon)$, $n^H(m=\{2,...,6\},4-\epsilon)$ \cite{KOMPANIETS2020}. We observe that all our estimates are compatible with Pad\'e-BL and BCM predictions from six-loop expansion \cite{KOMPANIETS2020}, five-loop expansion \cite{fiveloopomon} and pseudo-$\epsilon$ expansion \cite{fiveloopomon} (Table 2). Further it is interesting to compute critical exponents of chiral universality class for $m=2$ with $n\geq6$ since we obtain $n^+(2,3)\approx6$. We compute $\nu$, $\eta$ and $\gamma$ \cite{KOMPANIETS2020} using CE. CEF or CE-BL could not provide better accuracy. The series for $\eta$ takes the form $\epsilon^2 Q(\epsilon)$ which we handle similar to the way of $\omega$. We compute correction-to-scaling exponents $\omega_1$ and $\omega_2$ \cite{KOMPANIETS2020} using CE. $\omega_1$ and $\omega_2$ are interesting in physical point of view since they determine the stability of fixed points in RG flows. We tabulate all the estimates for chiral critical exponents in Table 3 to compare them with recent predictions from Pad\'e based resummation of six-loop expansion \cite{KOMPANIETS2020} and five-loop expansion \cite{fiveloopomon}. We observe that all our estimates have improved precision compared to the Pad\'e-BL and BCM predictions except for values of $n^-(6,3)$ and $\omega_2$ for $n=32$.
\begin{figure}[ht]
\centering
\begin{subfigure}{0.4\textwidth}
\includegraphics[width=1\linewidth, height=5cm]{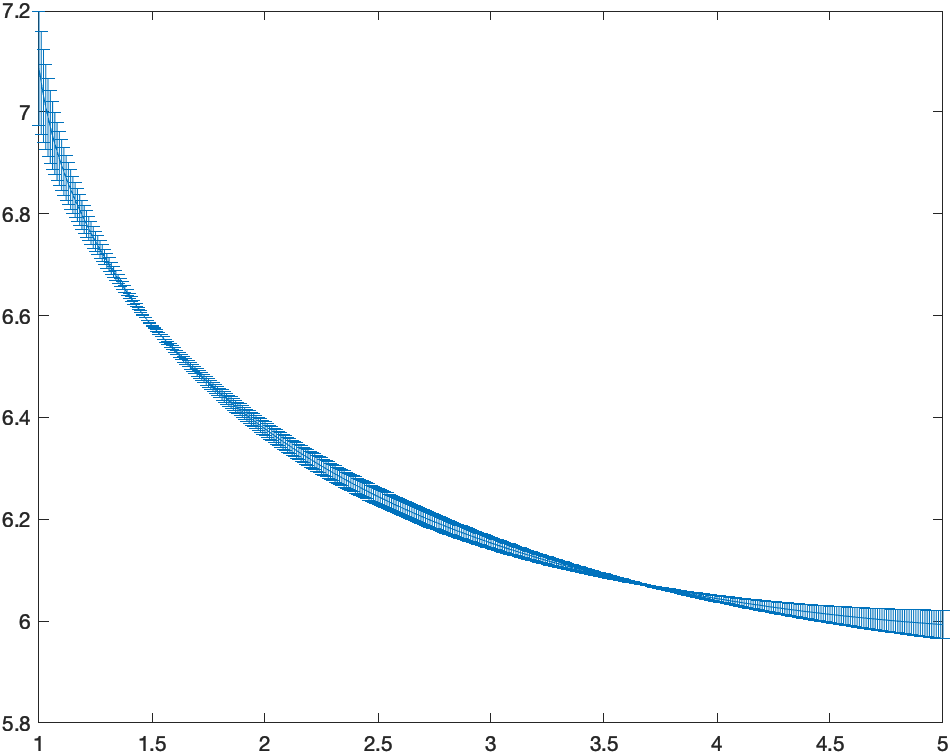} 
\caption{$n^+(2,3)$ vs $l$ for $l \in [1,5]$}

\end{subfigure}
\begin{subfigure}{0.4\textwidth}
\includegraphics[width=1\linewidth, height=5cm]{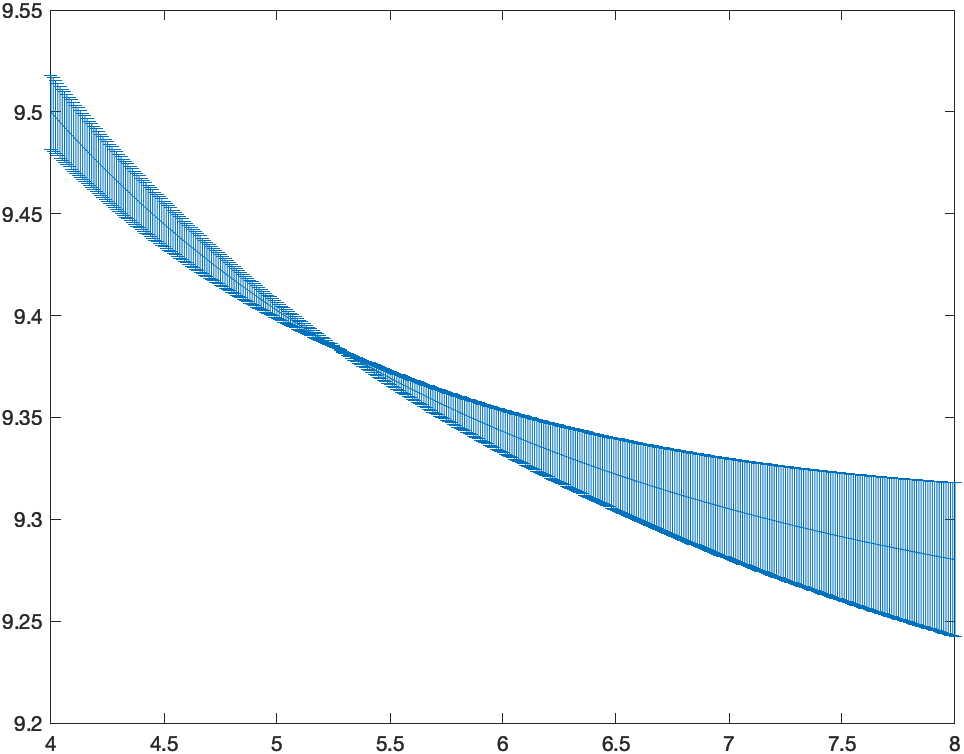}
\caption{$n^+(3,3)$ vs $l$ for $l \in [4,8]$}

\end{subfigure}

\caption{We plot the value of upper marginal order parameter $n^+$ derived from CE-BL estimate $S^2(D_3)$ vs shift parameter $l$, with the error bars showing the value of $(|S(D_{4}) - S(D_{3})| + |S(D_{4}) - S^2(D_{3})|)/2$.}

\end{figure}
\begingroup
\setlength{\tabcolsep}{2.75pt} 
\renewcommand{\arraystretch}{1} 
\begin{table}[htp]
\scriptsize
\begin{center}
\caption{Marginal dimensionalities $n^+$, $n^-$ and $n^H$ for class of $O(n)\times O(m)$ symmetric model for physically relevant three dimensional systems. CE-BL estimates are denoted by superscript*.} 

 \begin{tabular}{|c c c c c c|}
 
 \hline
Critical parameter & $m=2$ & $m=3$ & $m=4$ & $m=5$  & $m=6$ \\ [0.5ex] 
 \hline\hline
 
     $n^+(m,3)$
   
   & \begin{tabular}{c c c c c}
   & 5.99(27) (CE) \\
        & 6.07* ($l=3.69$) \\
        & 6.22(12) \cite{fiveloopomon} \\
        & 6.1(6) \cite{fiveloopomon} \\
        & 5.96(19) \cite{KOMPANIETS2020}
   \end{tabular} 
   & \begin{tabular}{c c c c c}
        & 9.25(22) (CE)\\
        & 9.384* ($l=5.27$)  \\ 
        & 9.9(3) \cite{fiveloopomon}\\
        & 9.5(5) \cite{fiveloopomon} \\
        & 9.32(19) \cite{KOMPANIETS2020} 
        \end{tabular}
        
     & \begin{tabular}{c c c c c}
        & 12.19(21) (CE)\\
        & 12.363* ($l=6.44$) \\ 
        & 13.2(6) \cite{fiveloopomon}\\
        & 12.7(7) \cite{fiveloopomon} \\
        & 12.3(3) \cite{KOMPANIETS2020}
        \end{tabular}
        & \begin{tabular}{c c c c c}
        & 15.04(22) (CE)\\
        & 15.215* ($l=7.47$)\\ 
        & 16.3(1.3) \cite{fiveloopomon}\\
        & 15.7(1.0) \cite{fiveloopomon} \\
        & 15.0(3) \cite{KOMPANIETS2020} 
        \end{tabular}
        & \begin{tabular}{c c c c c}
        & 17.83(23) (CE)\\
        & 18.003* ($l=8.43$) \\ 
        & 18.0(5) (BCM) \cite{KOMPANIETS2020}\\
        & 17.8(3) \cite{KOMPANIETS2020} 
        \end{tabular} \\
    
 \hline
     $n^-(m,3)$
   & \begin{tabular}{c c c c}
    & 1.964(3) (CE) \\
        & 1.968(5)* ($l=19.25$) \\
        & 1.970(3) \cite{KOMPANIETS2020}
   \end{tabular} 
   & \begin{tabular}{c c}
        & 1.403 (CE) \\
        & 1.406(1)* ($l=6.55$) \\
        & 1.408(4) \cite{KOMPANIETS2020}
   \end{tabular}
   & \begin{tabular}{c c}
             & 1.182(2) (CE)  \\
             & 1.182* ($l=21.32$) \\
             & 1.182(6) \cite{KOMPANIETS2020}
        \end{tabular} 
         & \begin{tabular}{c c}
             & 1.082(61) (CE)  \\
             & -- \\
             & 1.089(9) \cite{KOMPANIETS2020}
        \end{tabular}
         & \begin{tabular}{c c}
             & 1.325(5) (CE)  \\
             & -- \\
             & 1.066(12) \cite{KOMPANIETS2020}
        \end{tabular} \\
 \hline
    $n^H(m,3)$
    & \begin{tabular}{c c c c}
    & 1.478(13) (CE) \\
        & 1.459* ($l=0.6$) \\
        & 1.462(13) \cite{KOMPANIETS2020}
   \end{tabular} 
   & \begin{tabular}{c c}
        & 0.985(9) (CE) \\
        & 0.972* ($l=0.61$) \\
        & 0.973(11) \cite{KOMPANIETS2020}
   \end{tabular}
   & \begin{tabular}{c c}
             & 0.739(6) (CE)  \\
             & 0.729* ($l=0.6$) \\
             & 0.733(10) \cite{KOMPANIETS2020}
        \end{tabular} 
         & \begin{tabular}{c c}
             & 0.591(5) (CE)  \\
             & 0.583* ($l=0.61$) \\
             & 0.587(8) \cite{KOMPANIETS2020}
        \end{tabular}
         & \begin{tabular}{c c}
             & 0.493(4) (CE)  \\
             & 0.486* ($l=0.6$) \\
             & 0.488(7) \cite{KOMPANIETS2020}
        \end{tabular} \\
 \hline
     
\end{tabular}
\label{table 2}
\end{center}
\end{table}
\begingroup
\setlength{\tabcolsep}{2.75pt} 
\renewcommand{\arraystretch}{1} 
\begin{table}[htp]
\scriptsize
\begin{center}
\caption{Critical exponents $\nu$, $\eta$, $\gamma$, $\omega_1$ and $\omega_2$ for $m=2$ chiral universality class in  physically relevant three dimensional systems with noncollinear but coplanar ordering.} 

 \begin{tabular}{|c c c c c c|}
 
 \hline
Critical exponent & $n=6$ & $n=7$ & $n=8$ & $n=16$  & $n=32$ \\ [0.5ex] 
 \hline\hline
 
     $\nu$
   
   & \begin{tabular}{c c c}
   & 0.6516(21) (CE) \\
        & 0.65(2) \cite{KOMPANIETS2020}
   \end{tabular} 
   & \begin{tabular}{c c c c c}
        & 0.7117(5) (CE)\\
        & 0.71(4) \cite{fiveloopomon} \\
        & 0.713(8) \cite{KOMPANIETS2020} 
        \end{tabular}
        
     & \begin{tabular}{c c c c c}
        & 0.73398 (CE)\\
        & 0.75(4) \cite{fiveloopomon} \\
        & 0.745(11) \cite{KOMPANIETS2020} 
        \end{tabular}
        & \begin{tabular}{c c c c c}
        & 0.899(10) (CE)\\ 
        & 0.89(4) \cite{fiveloopomon}\\
        & 0.850(16)? \cite{KOMPANIETS2020} 
        \end{tabular} & \begin{tabular}{c c c c c}
        & 0.9368(33) (CE)\\
        & 0.94(2) \cite{fiveloopomon}\\
        & 0.940(17) \cite{KOMPANIETS2020} 
        \end{tabular} \\
    
 \hline
     $\eta$
   & \begin{tabular}{c c c c}
    & 0.04716(36) (CE) \\
        & 0.047(3) \cite{KOMPANIETS2020}
   \end{tabular} 
   & \begin{tabular}{c c}
        & 0.04436(69) (CE) \\
        & 0.045(3) \cite{KOMPANIETS2020}
   \end{tabular}
   & \begin{tabular}{c c}
             & 0.04144(99) (CE)  \\
             & 0.042(2) \cite{KOMPANIETS2020}
        \end{tabular} 
         & \begin{tabular}{c c}
             & 0.026156(29) (CE)  \\
             & 0.0261(7) \cite{KOMPANIETS2020}
        \end{tabular}
         & \begin{tabular}{c c}
             & 0.01609 (CE)  \\
             & 0.014(3) \cite{KOMPANIETS2020}
        \end{tabular} \\
 \hline
    $\gamma$
    & \begin{tabular}{c c c c}
    & 1.29561 (CE) \\
        & 1.27(3) \cite{KOMPANIETS2020}
   \end{tabular} 
   & \begin{tabular}{c c}
        & 1.39603(40) (CE) \\
        & 1.39(6) \cite{fiveloopomon} \\
        & 1.396(14) \cite{KOMPANIETS2020}
   \end{tabular}
   & \begin{tabular}{c c}
             & 1.45054 (CE)  \\
             & 1.45(6) \cite{fiveloopomon} \\
             & 1.461(17) \cite{KOMPANIETS2020}
        \end{tabular} 
         & \begin{tabular}{c c}
             & 1.795(35) (CE)  \\
             & 1.75(4) \cite{fiveloopomon} \\
             & 1.70(5)? \cite{KOMPANIETS2020}
        \end{tabular}
         & \begin{tabular}{c c}
             & 1.8565(55) (CE)  \\
             & 1.87(4) \cite{fiveloopomon} \\
             & 1.87(4) \cite{KOMPANIETS2020}
        \end{tabular} \\
 \hline
 $\omega_1$
   
   & \begin{tabular}{c c c}
   & 0.6516(21) (CE) \\
        & 0.65(2) \cite{KOMPANIETS2020}
   \end{tabular} 
   & \begin{tabular}{c c c c c}
        & 0.7117(5) (CE)\\
        & 0.71(4) \cite{fiveloopomon} \\
        & 0.713(8) \cite{KOMPANIETS2020} 
        \end{tabular}
        
     & \begin{tabular}{c c c c c}
        & 0.73398 (CE)\\
        & 0.75(4) \cite{fiveloopomon} \\
        & 0.745(11) \cite{KOMPANIETS2020} 
        \end{tabular}
        & \begin{tabular}{c c c c c}
        & 0.899(10) (CE)\\ 
        & 0.89(4) \cite{fiveloopomon}\\
        & 0.850(16)? \cite{KOMPANIETS2020} 
        \end{tabular} & \begin{tabular}{c c c c c}
        & 0.9368(33) (CE)\\
        & 0.94(2) \cite{fiveloopomon}\\
        & 0.940(17) \cite{KOMPANIETS2020} 
        \end{tabular} \\
        \hline
        $\omega_2$
   
   & \begin{tabular}{c c c}
   & 0.07327(42) (CE) \\ & [0.071,0.167] (BCM) \cite{KOMPANIETS2020} \\
        & [0.069,0.171] \cite{KOMPANIETS2020}
         
   \end{tabular} 
   & \begin{tabular}{c c c c c}
        & 0.3431(13) (CE)\\
        & 0.33(10) \cite{fiveloopomon} \\
        & 0.34(2) \cite{KOMPANIETS2020} 
        \end{tabular}
        
     & \begin{tabular}{c c c c c}
        & 0.4557(14) (CE)\\
        & 0.45(8) \cite{fiveloopomon} \\
        & 0.447(15) \cite{KOMPANIETS2020} 
        \end{tabular}
        & \begin{tabular}{c c c c c}
        & 0.77572(36) (CE)\\ 
        & 0.77(2) \cite{fiveloopomon}\\
        & 0.771(6) \cite{KOMPANIETS2020} 
        \end{tabular} & \begin{tabular}{c c c c c}
        & 0.891(27) (CE)\\
        & 0.90(1) \cite{fiveloopomon}\\
        & 0.904(8) \cite{KOMPANIETS2020} 
        \end{tabular} \\
    
 \hline
     
\end{tabular}
\label{table 3}
\end{center}
\end{table}
\section{Randomly diluted Ising model}
Randomly diluted Ising model (RIM) for systems with "frozen" impurities have different universality class of critical exponents derived from that of pure Ising model \cite{RIM1974,RIM1975}. Studying this model using RG approach is interesting since other disordered systems such as randomly site-diluted Ising model \cite{RSIM1998,RSIM2003}, randomly bond-diluted Ising model \cite{RBIM2004,RBIM2010} and $\pm J$ Ising model \cite{PMJ2007,PMJ2013} fall under the same category of RIM universality class. Experimentally the critical behaviours are studied in crystalline mixture of two compounds such as ordering in an anisotropic uniaxial antiferromagnet of $Fe F_2$ or $Mn F_2$ with $Zn F_2$ as impurity. \\ Initial RG studies on RIM model has led to conclusion that critical exponents have to be in the powers of $\sqrt{\epsilon}$ instead of $\epsilon$ \cite{grinstein1976,khmelnitskii1975s}. However the most recent studies show that resummation of critical exponents in $\sqrt{\epsilon}$ expansion do not give reliable estimates in comparison with predictions from other theoretical approaches \cite{folk2000,RIM2021}. For instance the recent six-loop $\sqrt{\epsilon}$ expansions derived for $\nu$ and $\gamma$ are in the form of \cite{RIM2021}
\begin{equation}
    \nu_{\sqrt{\epsilon}} = 0.5+0.0841158 \epsilon^{1/2}-0.016632\epsilon+0.0477535\epsilon^{3/2}+0.272584\epsilon^2+0.223298\epsilon^{5/2}+ O(\epsilon^3) 
\end{equation}
and 
\begin{equation}
    \gamma_{\sqrt{\epsilon}} = 1+0.168232\epsilon^{1/2}-0.0285471\epsilon+0.0788288\epsilon^{3/2}+0.564505\epsilon^2+0.440615\epsilon^{5/2}+ O(\epsilon^3).
\end{equation}
We construct CEF sequences such as in Eq. (5) for variable $t$ with a change in variable $\epsilon \rightarrow \epsilon t^2$ for $d=3$ and finally equate $t$ to unity. We get CEF estimates $\nu_{\sqrt{\epsilon}} = 0.57956(16)$ and $\gamma_{\sqrt{\epsilon}} = 1.16108(19)$ which are more precise than the recent Pad\'e-BL predictions $\nu_{\sqrt{\epsilon}} = 0.577(31)$ and $\gamma_{\sqrt{\epsilon}} = 1.172(55)$ \cite{RIM2021}. However we come to the same conclusion that these values of critical exponents obtained from resummation of $\sqrt{\epsilon}$ expansions are in contradiction from theoretical and experimental results, further confirming that these series are not resummable. One can further try resummation of the RG functions directly through continued functions instead of $\sqrt{\epsilon}$ expansions to obtain critical exponents as tried previously \cite{folk2000,RIM2021}. \\
\section{Conclusion}
Simple methods using continued functions were utilised to derive precise critical parameters and critical exponents from six-loop RG perturbative expansions of $n$-vector model with cubic anisotropy, $O(n)\times O(m)$ spin models and weakly disordered Ising model. This is useful in better defining the nature of continuous phase transitions in their corresponding physical systems. To summarise there is an interesting quote in the recent work of Kompaniets and Panzer regarding resummation methods \cite{six-O(n)} : {\it In work on resummation, there is always an undeclared parameter: the number of methods tried and rejected before the paper was written,} which was followed by their comment "We stopped counting". Taking into consideration the essence of this problem, we have devised resummation methods with ease of use and less computations, which provide estimates with better accuracy than Pad\'e based methods. Also Pad\'e methods are typically riddled with spurious poles, where a thorough inspection is required to remove them when finding a reliable estimate. However in our approach one has to further rigorously study which strategic method using CE, CEF or CE-BL provides best accuracy and reliable convergence based on the numerical structure of the perturbation series. Also in all our cases we have handled only physically relevant systems with $d=3 (\epsilon=1$), further trying to study $d=2 (\epsilon=2)$ systems can cause trivial problems to emerge from the structure of continued functions we have considered. Such problems can perhaps be removed by implementing long-order asymptotic behaviour of the coefficients in the perturbation series if available. 
\bibliographystyle{ieeetr}
\bibliography{sample.bib}
\end{document}